\documentclass[preprint,showpacs,english,aps]{revtex4-1}
\usepackage{babel,rotating,dcolumn}
\usepackage{colordvi,graphicx,color,amsbsy,amsmath,bm}
\begin{document}
\title[Examination of the four-fifths law for incompressible magnetohydrodynamic turbulence]
{Examination of the four-fifths law for longitudinal third-order moments in incompressible magnetohydrodynamic turbulence in a periodic box}
\author{Katsunori Yoshimatsu} 
\email{yoshimatsu@nagoya-u.jp}
\affiliation{Department of Computational Science and Engineering, Nagoya University, Nagoya, 464-8603, Japan}
\date{\today}
\begin{abstract}
The four-fifths law for third-order longitudinal moments is examined, 
by the use of direct numerical simulation data on three-dimensional forced incompressible magnetohydrodynamic (MHD) turbulence without a uniformly imposed magnetic field in a periodic box.
The magnetic Prandtl number is set to one, 
and the number of grid points is $512^3$.
A generalized K\'arm\'an-Howarth-Kolmogorov equation for second-order velocity moments in isotropic MHD turbulence is extended to anisotropic MHD turbulence by means of a spherical average over the direction of ${\bm r}$.
Here, ${\bm r}$ is a separation vector.
The viscous, forcing, anisotropy and nonstationary terms in the generalized equation are quantified.
It is found that the influence of the anisotropic terms on the four-fifths law is negligible at small scales, compared to that of the viscous term. 
However, the influence of the directional anisotropy, which is measured by 
the departure of the third-order moments in a particular direction of ${\bm r}$ from the spherically averaged ones,
on the four-fifths law is suggested to be substantial, at least in the case studied here.
\end{abstract}
\maketitle
\section{Introduction}
Magnetohydrodynamic (MHD) turbulence, as well as hydrodynamic (HD) turbulence, is ubiquitous in nature.
Turbulence is characterized by a large number of degrees of freedom, a wide range of active scales, and strong nonlinearity.
It is thought that there is a certain kind of statistical universality at sufficiently small scales in fully developed turbulence away from boundaries (see e.g. Refs. \cite{Frisch, Biskamp}).
A striking feature of this universality is the existence of exact statistical laws for homogeneous incompressible turbulence at sufficiently high Reynolds numbers.
Such exact laws are rare. 

In three-dimensional (3D) incompressible homogeneous isotropic HD turbulence,
we have Kolmogorov's 4/5 law \cite{K41},
\begin{equation}
\langle \{ \delta u_L ({\bm x},{\bm r},t) \}^3 \rangle = - \frac{4}{5} {\bar \epsilon} r, \label{HD4/5}
\end{equation}
which can be derived from the K\'arm\'an-Howarth-Kolmogorov equation (hereafter referred to as the KHK equation) \cite{KH}.
Here, $\delta u_L({\bm x},{\bm r},t)$ is the longitudinal velocity difference between the points at ${\bm x} + {\bm r}$ and ${\bm x}$,
$\langle \cdots \rangle$ denotes the ensemble average,
${\bar \epsilon}$ is the mean energy dissipation rate per unit mass, $r=|{\bm r}|$, and $t$ is time.
For homogeneous turbulence,
the average $\langle \xi({\bm x},{\bm r},t) \rangle$ is independent of the position ${\bm x}$,
where $\xi({\bm x},{\bm r},t)$ is any quantity obtained from any of the field quantities at positions ${\bm x}$ and ${\bm x}+{\bm r}$.
This 4/5 law is exact in the range $\eta_{\mathrm{K}} \ll r \ll L$ at infinitely large Reynolds number, 
where $L$ and $\eta_{\mathrm{K}}$ are the characteristic length scales of the energy-containing range and the Kolmogorov microscale, respectively.
However, any real turbulence, in which the Reynolds number and the scale range are finite, is not statistically isotropic, homogeneous, or stationary in a strict sense, owing to the influences of viscosity, external forcing, large-scale anisotropy, and so on.
Since the law is exact, extensive studies on the law have been done experimentally, theoretically and numerically, 
in order to gain a quantitative understanding of the universality (see e.g. a recent review \cite{Ishihara}).

The 4/5 law, Eq. (\ref{HD4/5}), is extended to 3D incompressible MHD turbulence
by using a generalized KHK equation for the second-order velocity moments under the assumption that the flow is homogeneous and isotropic \cite{Chandra,YRS}:
\begin{equation}
\langle \{ \delta u_L ({\bm x},{\bm r},t) \}^3 \rangle - 6 \langle b_L^2({\bm x},{\bm r},t) \delta u_L({\bm x},{\bm r},t) \rangle = - \frac{4}{5} {\bar \epsilon_t} r, \label{MHD4/5}
\end{equation}
where $b_L$ is the longitudinal component of the magnetic field normalized by 
$(4\pi \rho)^{1/2}$, ${\bar \epsilon_t} $ is the total mean energy dissipation rate per unit mass,
and $\rho$ is the density of the conducting fluid.
This law is exact in the range $\eta_{\mathrm{IK}} \ll r \ll L$ for infinitely large kinetic and magnetic Reynolds numbers, 
where $\eta_{\mathrm{IK}}$ is the Iroshnikov and Kraichnan microscale.
In Ref. \cite{Chandra}, another exact law for the third-order moments originated from the induction equations was derived
on the basis of the generalized KHK equation for the second-order moments of the magnetic field in homogeneous isotropic MHD turbulence.

Other exact laws were also derived; namely, the three divergence laws for third-order moments of incompressible homogeneous anisotropic MHD turbulence in the range $\eta_{\mathrm{IK}} \ll r \ll L$, 
which are equivalent to the scale-independence of the fluxes of total energy, cross helicity, and magnetic helicity \cite{Podesta2008}.
The laws for total energy and cross helicity correspond to the $4/d$ laws for the third-order mixed structure functions of the Els\"asser variables in $d$-dimensional incompressible homogeneous MHD turbulence, 
which were first derived in Ref. \cite{PPYag} under the assumption of flow isotropy.
These 4/3 laws $(d=3)$ are used in space craft measurements.
The presence of an energy cascade in the solar wind was shown \cite{SV}, and
the inertial range cascade rate was also directly determined \cite{Osman}.
Furthermore, the $4/d$ laws have been extended to homogeneous incompressible MHD turbulence with constant shear \cite{Wan2009}.
Data analysis using the direct numerical simulation (DNS) in two dimensions showed that 
the shear effect broadens the circumstances under which the law can be applied \cite{Wan2010}.
The $4/3$ law for total energy has also been generalized to incompressible homogeneous isotropic Hall MHD turbulence \cite{GaltierH}.

In this paper, we examine the 4/5 law, Eq. (\ref{MHD4/5}), for 3D incompressible MHD turbulence in the absence of a uniformly imposed magnetic field in a periodic box.
The average $\langle \cdots \rangle $ is independent of ${\bm x}$ for turbulence under periodic boundary conditions,
if the average is understood as the volume average over the fundamental periodic domain. 
Owing to external forcing, its large-scale flow anisotropy, and the finite scale range at finite kinetic and magnetic Reynolds numbers, 
the flow is not perfectly isotropic or stationary.
The mean magnetic field obtained by averaging over an appropriate volume in the periodic box cannot be removed from the system by a Galilean transformation, 
in contrast to the mean velocity in the volume considered.
Therefore, a large-scale magnetic field may play a substantial role in small-scale anisotropy.
The term $\langle b_L^2 \delta u_L \rangle$ in Eq. (\ref{MHD4/5}), which cannot be represented only by $\delta b_L$ and $\delta u_L$, includes large-scale information about $b_L$.
In Refs. \cite{MMDM} and \cite{MP2007}, it was shown that
the large-scale magnetic field leads to local anisotropy at smaller scales for  MHD turbulence without a uniformly imposed external magnetic field.
The influence of large-scale anisotropy can also be brought by nonlinear interactions nonlocal in scale.
It was shown that the nonlinear interactions in MHD turbulence are significantly more nonlocal than in HD turbulence \cite{Doma,Teaca}. 
Readers interested in these interactions may refer to a recent review \cite{MiniAnnu}.

In the next section, 
we extend the generalized KHK equation for isotropic MHD turbulence to anisotropic MHD turbulence,
in order to get quantitative insight into the influences of anisotropy, as well as large-scale forcing, viscosity, and non-stationarity, on the 4/5 law.
In Sec.~III, we describe DNS of 3D incompressible forced MHD turbulence without a uniformly imposed magnetic field in a periodic box.
The forcing is imposed only on the large-scale velocity field.
Section IV presents a numerical examination of the 4/5 law based on the generalized equation.
Emphasis is placed on the influence of anisotropy.
Finally, conclusions are drawn in Sec.~V. 
\section{Formulation}
\subsection{Basic Equations}
We consider the 3D MHD motion of incompressible conductive fluid of density $\rho$ in the Cartesian coordinates ${\bm x}=(x_1,x_2,x_3)$.
The motion obeys the following governing equations:
\begin{eqnarray}
& & \partial_t u_i + u_j \partial_j u_i = b_j \partial_j b_i - \partial_i P + \nu \partial_j \partial_j u_i + f_i^u, \label{NS}\\
& & \partial_t b_i + u_j \partial_j b_i = b_j \partial_j u_i + \eta \partial_j \partial_j b_i + f_i^b,  \label{induction} \\
& & \partial_j u_j =0, \label{divu} \\
& & \partial_j b_j =0, \label{divb}
\end{eqnarray}
in a periodic box with sides of length $2\pi$.
Here, $u_i({\bm x},t)$ is the $i$th component of the velocity field,
$b_i({\bm x},t)$ is the $i$th component of the magnetic field normalized by $(4\pi \rho)^{1/2}$, 
$f^u_i({\bm x},t)$ and $f^b_i({\bm x},t) $ are the $i$th components of the solenoidal external forces, 
$P({\bm x},t)$ is the total pressure, which is normalized by $\rho$, including the magnetic pressure, 
$\nu$ is the kinematic viscosity, $\eta$ is the magnetic diffusivity,
$\partial_t=\partial/\partial t$, and $\partial_j=\partial/\partial x_j$. 
The summation convention over $\{1,2,3\}$ is used for repeated alphabetical subscripts, excluding $t$.
The arguments $t$ and ${\bm x}$ are omitted, unless otherwise stated.
The average $\langle \cdots \rangle $ hereafter denotes the volume average over the periodic box. 

Multiplying $u_i$ and $b_i$ with Eqs. (\ref{NS}) and (\ref{induction}), respectively, and taking the volume average, 
we obtain the evolution equations for the kinetic energy ${\bar E_u}$ and magnetic energy ${\bar E_b}$;
\begin{eqnarray}
\frac{{d}{\bar E_u}}{{d}t} &=& 
-{\bar \epsilon_u} + \langle u_i f^u_i \rangle, \label{eneu} \\
\frac{{d}{\bar E_b}}{{d}t} &=& -{\bar \epsilon_b}
+\langle b_i f^b_i \rangle, \label{eneb}
\end{eqnarray}
where ${\bar E_u}= \langle u_i u_i \rangle/2$, ${\bar E_b}= \langle b_i b_i \rangle/2$, and
\begin{eqnarray}
{\bar \epsilon}_u &=& {\bar \epsilon}_\nu + \langle b_i b_j \partial_j u_i \rangle , \label{epu} \\
{\bar \epsilon}_b &=& {\bar \epsilon}_\eta -\langle b_i b_j \partial_j u_i \rangle . \label{epb}
\end{eqnarray}
Here, 
${\bar \epsilon}_\nu = \nu \langle \partial_j u_i \partial_j u_i \rangle $ and ${\bar \epsilon}_\eta = \eta \langle \partial_j b_i \partial_j b_i \rangle $.
The total mean energy dissipation rate per unit mass ${\bar \epsilon_t}$ is given by ${\bar \epsilon_t} = {\bar \epsilon_\nu} + {\bar \epsilon_\eta}$.
By the use of Eqs. (\ref{eneb}) and (\ref{epb}), Eq. (\ref{epu}) can be rewritten as
\begin{equation}
{\bar \epsilon}_u = {\bar \epsilon}_t - \langle b_i f^b_i \rangle + \frac{{d}{\bar E_b}}{{d}t} . \label{epub}
\end{equation}
\subsection{Generalization of K\'arm\'an-Howarth-Kolmogorov equation for anisotropic MHD turbulence}
In this subsection, we derive a generalized KHK equation for homogeneous anisotropic MHD turbulence.
The equation describes the time evolutions of the second-order velocity moments.
The generalization procedure follows that in Ref. \cite{Kaneda2008}, in which a generalized KHK equation for anisotropic HD turbulence was derived to examine the 4/5 law, Eq. (\ref{HD4/5}).

We start with the Navier-Stokes equations, together with the solenoidal conditions for the velocity and magnetic fields at points ${\bm x}$ and ${\bm x}'(\equiv {\bm x} + {\bm r})$, 
where the point ${\bm x}'$ is independent of ${\bm x}$.
The equations and the conditions at point ${\bm x}$ are given by Eqs. (\ref{NS}), (\ref{divu}) and (\ref{divb}), 
while those at point ${\bm x}'$ are obtained by replacing $u_i({\bm x})$, $b_i({\bm x})$, $P({\bm x})$, and $\partial_i$ with $u_i'$, $b_i'$, $P'$, and $\partial'_i$, respectively, in Eqs. (\ref{NS}), (\ref{divu}), and (\ref{divb}).
Here $u_i'=u_i({\bm x}')$, $b_i'=b_i({\bm x}')$, $P'=P({\bm x}')$ and  $\partial'_i=\partial/\partial x_i'$.

Subtraction of Eq. (\ref{NS}) at point ${\bm x}$ from the Navier-Stokes equations at ${\bm x}'$ results in the equations for the velocity increment $\delta u_i({\bm x},{\bm r},t)$ defined as $\delta u_i= u_i({\bm x}',t)-u_i({\bm x},t)$.
Multiplying the resulting equations by $2\delta u_i$, applying contraction with respect to $i$, and averaging over the periodic domain, we obtain
\begin{eqnarray}
\partial_t D_{ii} 
 &=& - \frac{\partial}{\partial r_j} \langle  (\delta u_i \delta u_i) \delta u_j \rangle ({\bm r}) + 2 \nu \frac{\partial}{\partial r_j}\frac{\partial}{\partial r_j} D_{ii} -4 {\bar \epsilon_u } \nonumber \\
 &+&  2 \frac{\partial }{\partial r_j} \left\{
\left\langle b_j b_i \delta u_i \right \rangle ({\bm r}) + \left\langle b'_j b'_i  \delta u_i \right \rangle ({\bm r}) \right\} 
+ 2 \langle \delta u_i \delta f_i^u\rangle ({\bm r}), \label{Dii}
\end{eqnarray}
where $D_{ii} ({\bm r})=\langle  \delta u_i \delta u_i \rangle $ and $\delta f_i^u= f_i^u({\bm x}')-f_i^u({\bm x})$.
The terms in Eq. (\ref{Dii}), except for the constant $4{\bar \epsilon}_u$, depend only ${\bm r}$ and $t$,
and are independent of ${\bm x}$.
In arriving at Eq. (\ref{Dii}), we have used the solenoidal conditions, $\partial_j=-\partial/\partial r_j$, $\partial_j'=\partial/\partial r_j$, Eq. (\ref{epu}), and the following relations:
\begin{eqnarray}
& & 2 \nu \langle \delta u_i \partial_j \partial_j \delta u_i \rangle = \nu \frac{\partial}{\partial r_j} \frac{\partial}{\partial r_j} D_{ii}  -2 {\bar \epsilon}_\nu, \\
& & \left\langle \delta u_i \left\{ \partial'_j (b'_j b'_i) - \partial_j (b_j b_i) \right\} \right\rangle = -2 ({\bar \epsilon}_u - {\bar \epsilon}_{\nu}) + \frac{\partial}{\partial r_j}  \left( 
\langle b_j b_i \delta u_i  \rangle + \langle b'_j b'_i \delta u_i \rangle \right), \\
& & \langle \delta u_i (\partial_i + \partial_{i}') \delta P \rangle =0, \label{pre}
\end{eqnarray}
where $\delta P= P({\bm x}')-P({\bm x})$.
Relation (\ref{pre}) was shown for HD turbulence in Ref. \cite{Antonia}.

Let $\langle \xi({\bm r}) \rangle_r$ be the average of $\xi({\bm r})$ over ${\bm r}$ on the spherical surface of radius $r=|{\bm r}|$ with a center at ${\bm r}={\bm 0}$; that is,
\begin{equation}
\left\langle \xi({\bm r}) \right\rangle_r \equiv 
\frac{1}{4\pi r^2} \int \!\!\! \int_{|{\tilde {\bm r}}|=r} \xi({\tilde {\bm r}}) {d} \Omega,
\end{equation}
where $\int \!\! \int_{|{\bm r}|=r} {d} \Omega$ denotes the integral over the  spherical surface.
Note that $\langle \xi (-{\bm r})\rangle_r = \langle \xi ({\bm r})\rangle_r$; that is, 
$\langle \xi ({\bm r})\rangle_r$ has reflection symmetry with respect to ${\bm r}$ for any $\xi$.
The divergence theorem yields
\begin{equation}
\int \!\!\! \int \!\!\! \int_{|{\tilde {\bm r}}|\le r}    
\frac{\partial}{\partial {\tilde r_j}} \langle \zeta_j \rangle ({\tilde {\bm r}}) {{d}}^3 {\tilde {\bm r}} = \int \!\!\! \int_{|{\tilde {\bm r}}|=r} \langle \zeta_j \rangle ({\tilde {\bm r}})
\frac{{\tilde r_j}}{{\tilde r}} {{d}} \Omega 
= 4 \pi r^2 \langle \langle \zeta_L \rangle \rangle_r,
\end{equation}
where $\zeta_L = \zeta_j r_j/r$, 
and $\zeta_j$ denotes the $j$th component of any vector function with respect to ${\bm x}$, ${\bm r}$ and $t$. 

Then, integrating Eq. (\ref{Dii}) over the inside of the sphere of radius $r$ with a center at ${\bm r}={\bm 0}$, 
using $\langle b'_j b'_i \delta u_i \rangle ({\bm r}) = - \langle b_j b_i \delta u_i \rangle (-{\bm r}) $, 
and applying the divergence theorem, we have
\begin{eqnarray}
\left\langle \langle \delta u_i \delta u_i \delta u_L \rangle \right\rangle_r 
- 4 \left\langle \langle b_Lb_i \delta u_i \rangle \right\rangle_r = - \frac{4}{3} {\bar \epsilon}_u  r + H_t (r) + H_f(r) + H_{\nu}(r), \label{uYag}
\end{eqnarray}
where
\begin{eqnarray}
H_t(r) &=& -\frac{1}{4\pi r^2 } \int \!\!\! \int \!\!\! \int_{|{\tilde {\bm r}}| \le r} \partial_t D_{ii} ({\tilde {\bm r}})  {d}^3 {\tilde {\bm r}}, \\
H_f(r) &=& \frac{1} {2 \pi r^2 } \int \!\!\! \int \!\!\! \int_{|{\tilde {\bm r}}| \le r} 
\langle \delta f_i^u ({\tilde {\bm r}}) \delta u_i ({\tilde {\bm r}})\rangle {d}^3 {\tilde {\bm r}}, \\
H_{\nu}(r) &=& 2 \nu \frac{\partial}{\partial r} \langle  D_{ii}({\bm r}) \rangle_r.
\end{eqnarray}
Next, let $H_u(r)$ and $H_b(r)$ be respectively defined by 
\begin{eqnarray}
H_u(r) &=& \frac{1}{3r^3} \frac{\partial}{\partial r} \left( r^4 \langle\langle (\delta u_L)^3 \rangle\rangle_r \right) - \langle\langle \delta u_i \delta u_i \delta u_L \rangle\rangle_r, \label{aHu} \\
H_b(r) &=& - 4 \left\{ \frac{1}{2r^3} \frac{\partial}{\partial r} 
\left( r^4 \langle\langle b_L^2 \delta u_L \rangle\rangle_r \right)
- \langle\langle b_L b_i \delta u_i \rangle\rangle_r \right\}. \label{aHb}
\end{eqnarray}
Then, a little algebra after subtraction of Eqs. (\ref{aHu}) and (\ref{aHb}) from Eq. (\ref{uYag}) results in
\begin{equation}
\langle\langle (\delta u_L)^3  \rangle\rangle_r- 6 \langle\langle b_L^2 \delta u_L \rangle\rangle_r = - \frac{4}{5} {\bar \epsilon}_u r + I_t (r) + I_f(r) + I_{\nu}(r) + I_u (r)+ I_b (r), \label{MHDKHu}
\end{equation}
where
\begin{equation}
I_{\alpha} = \frac{3}{r^4} \int_0^r {\tilde r}^3 H_{\alpha} ({\tilde r}) {d} {\tilde r} \quad (\alpha=t,f,\nu,u,b). \label{Ialpha}
\end{equation}
Here we call Eq. (\ref{MHDKHu}) the generalized KHK equation for anisotropic MHD turbulence.
The third-order moments $-6\langle\langle b_L^2 \delta u_L \rangle\rangle_r$ and $\langle\langle (\delta u_L)^3  \rangle\rangle_r$ have a direct connection with the fluxes of the kinetic energy, 
because the moments come from the nonlinear terms $b_j\partial_j b_i$ and $u_j\partial_j u_i$ in Eq. (\ref{NS}). 
In deriving Eq. (\ref{MHDKHu}), we have not used Eq. (\ref{induction}).
Therefore, Eq. (\ref{MHDKHu}) holds, if the Hall effect or the forcing acts on the magnetic field.
Without a magnetic field, Eq. (\ref{MHDKHu}) reduces to the generalized KHK equation for homogeneous anisotropic HD turbulence given by Eq. (24) in Ref. \cite{Kaneda2008}.

Since any volume-averaged quantity $\langle \xi \rangle({\bm r})$ is independent of the direction of ${\bm r}$ for strictly isotropic flow, 
the following relations hold:
\begin{eqnarray}
\langle \delta u_L \delta u_i \delta u_i \rangle &=& \frac{1}{3r^3} \frac{\partial}{\partial r} \left\{r^4  \langle (\delta u_L)^3 \rangle  \right\}, \\
\langle b_L b_i \delta u_i \rangle &=& \frac{1}{2r^3} \frac{\partial}{\partial r} \left(r^4 \langle b_L^2 \delta u_L \rangle \right).
\end{eqnarray}
Therefore, $H_u=H_b=0$, and thus $I_u=I_b=0$.
Hereafter, we use $I_u$ and $I_b$ as anisotropic measures.

If all $I_{\alpha}$ $(\alpha=t,f,\nu,u,b)$ are negligible in the inertial subrange,
and if ${d} {\bar E}_b/{d} t =0$, 
then, using Eq. (\ref{epub}), we have
\begin{equation}
\langle\langle (\delta u_L)^3 \rangle\rangle_r - 6 \langle\langle b_L^2 \delta u_L \rangle\rangle_r = - \frac{4}{5} ({\bar \epsilon}_t - \langle b_i f^b_i \rangle) r. \label{g45}
\end{equation}
This is expressed by the use of only longitudinal third-order moments.
Equation (\ref{g45}) is consistent with the results in Refs. \cite{Chandra,YRS} in the case when $f_i^b=0$.
These works use a generalized KHK equation for homogeneous isotropic MHD turbulence and the evolution equation of mean kinetic and magnetic energies for the case that $f^b_i=0$, in arriving Eq. (\ref{MHD4/5}). 
Equation (\ref{MHD4/5}) is in accordance with Eq. (7) for 3D case of Ref. \cite{PPKH} for generalized KHK equations for homogeneous isotropic MHD turbulence based on the Els\"asser variables.
The accordance shows that the latter does not contain the third-order moments arising from the induction equations (\ref{induction}).

\section{Direct Numerical Simulation}
We performed DNS of 3D incompressible MHD turbulence in a periodic box with sides of length $2\pi$.
Neither a uniform magnetic field nor external forcing for the magnetic field is imposed.
The magnetic Prandtl number is set to $1$ (i.e., $\eta=\nu$),
and the number of grid points in each direction of the Cartesian coordinates, $N$, is $512$.
The total number of grid points is $512^3$.
Equations (\ref{NS})$-$(\ref{divb}) are computed by the use of a fourth-order Runge-Kutta method for time integration and a Fourier pseudo-spectral method.
The aliasing errors are removed by a phase shift method. 
Only modes with wave numbers satisfying $k < 2^{1/2}N/3$ are retained, 
where $k=|{\bm k}|$, and ${\bm k}$ is a wave vector.
The wave number increment is $1$, and the minimum wave number is $1$. 
The time increment is taken to be equal to $1.5 \times 10^{-3}$, and $\nu=\eta=3.3 \times 10^{-4}$. 
We imposed a solenoidal random force only on the velocity field in the wave number range $1 \le k< 2.5$.
The correlation time of the force and its intensity are set to $2.1$ and $0.9\times 10^{-3}$, respectively.
Readers interested in details of how to generate the random force are referred to the appendix of Ref. \cite{Yoshida}. 
\begin{table}[t!]
\begin{center}
\caption[smallcaption]{Turbulence characteristics at the final time $t=t_f$.}
\label{DNS_characteristics}
\vspace*{0.5cm}
\begin{tabular}{cc ccccccccc}
\hline\hline
${\bar E}_u$ & ${\bar E}_b$ & ${\bar \epsilon}_t$ & ${\bar \epsilon}_u$ & ${\bar \epsilon}_\nu$ & $L_u$ & $\eta_{\mathrm{IK}}  $ & $R^{u}_{\lambda}$ & $R^{b}_{\lambda}$ & ${\mathcal H}^C  $ & ${\mathcal H}^M$ \\
\hline
$ 0.233 $ & $ 0.604 $  & $0.114$ & $0.114$ & $0.0438$ & $0.934$  &  $8.45\times 10^{-3}$ & $158$ & $323$ & $5.61 \times 10^{-2}$ & $0.657 $  \\ 
\hline\hline
\end{tabular}
\end{center}
\end{table}

The initial flow, which is the same as that used in Ref. \cite{scaledependent}, is given by a linear superposition of a random flow and a deterministic one.
The initial kinetic energy ${\bar E_u}$ and magnetic energy ${\bar E_b}$ are set to ${\bar E_u}={\bar E_b}=0.5$, 
where ${\bar E_u}=\langle u_i u_i \rangle/2$ and ${\bar E_b}=\langle b_i b_i \rangle/2$.
The initial cross helicity ${\bar H^C}$, defined as ${\bar H^C}=\langle u_i b_i \rangle $, is almost zero: ${\bar H^C}=3.78 \times 10^{-2}$, and the magnetic helicity ${\bar H^M}$, defined as ${\bar H^M}=\langle a_i b_i \rangle$, is set to ${\bar H^M}=0.515$, 
where $a_i$ is the vector potential of $b_i$.
The relative cross helicity ${\mathcal H}^C$ and magnetic helicity ${\mathcal H}^M$ are given as ${\mathcal H}^C=3.78 \times 10^{-2}$ and ${\mathcal H}^M=0.691$, where ${\mathcal H}^C={\bar H^C}/\{2({\bar E}_u{\bar E}_b)^{1/2}\} $ and ${\mathcal H}^M={\bar H^M}/\{2({\bar E}_a{\bar E}_b)^{1/2}\} $.
Here ${\bar E}_a= \langle a_i a_i \rangle /2$.

\begin{figure}[t]
\begin{center}
\includegraphics[width=8cm,keepaspectratio]{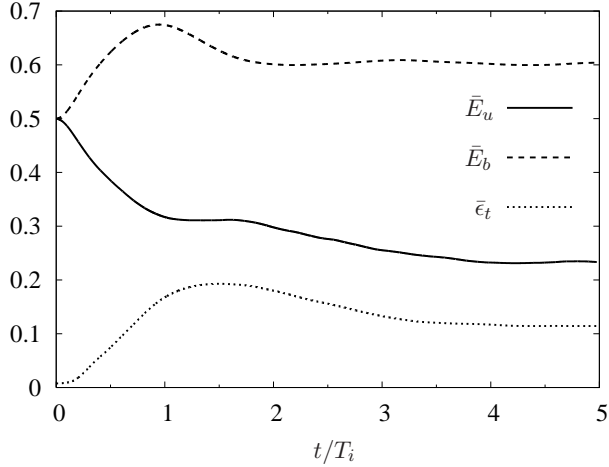}
\end{center}
\caption{Time evolution of 
${\bar E}_u$, 
${\bar E}_b$, and 
${\bar \epsilon}_t$.}
\label{ene_diss}
\end{figure}

The simulation is performed up to $t=t_f = 4.97 T_i$ 
until the flow becomes statistically quasi-stationary; that is,
${\bar E}_u$, ${\bar E}_b$ and the total energy dissipation rate per unit mass ${\bar \epsilon}_t$ remain almost constant, as shown in Fig. \ref{ene_diss}.
Here, $T_i$ is the initial large eddy turnover time defined by $T_i=L_u/u_0 =1.53$, 
$L_u$ is the integral length scale defined by $L_u ={\pi}/({2 u_0^2}) \int_0^{k_{\mathrm{max}}} {{d}} k \, {E_u(k)}/{k}$,
$u_{0}= \left( 2 {\bar E}_u / 3 \right)^{1/2}$,
$E_u(k)$ is the kinetic energy spectrum,
and ${k_{\mathrm{max}}}$ is the maximum wave number.
The value of $|{\bar H}^C|$ remains less than $4.22 \times 10^{-2}$, and ${\bar H}^M$ slowly decays to 0.506.
At $t=t_f$, the difference between ${\bar \epsilon}_u$ and ${\bar \epsilon}_t$ is less than $0.01 \%$ of ${\bar \epsilon}_t$, 
(i.e., ${\bar \epsilon}_u \simeq {\bar \epsilon}_t $), 
and the modulus of the time-derivative of ${\bar E}_b$ is very small [i.e., $|{{d}}{\bar E}_b/{{d}}t| \simeq 10^{-5}(\simeq 10^{-4}{\bar \epsilon}_t)$]. 
We did not take the time average here,
because a much longer large-eddy turnover time will be needed to achieve an averaged value of ${{d}} {\bar E}_b/{d} t $ that is sufficiently small.

\begin{figure}[tb]
\begin{center}
\includegraphics[width=8cm,keepaspectratio]{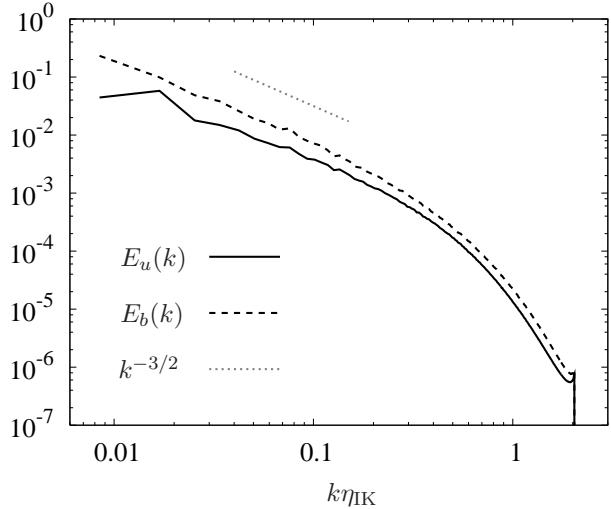}
\end{center}
\caption{
Kinetic and magnetic energy spectra, $E_u(k)$ and $E_b(k)$ vs nondimensional wave number $k \eta_{\mathrm{IK}}$. 
As a reference, the IK spectrum \cite{Iroshnikov,Kraichnan} (i.e., $k^{-3/2}$) is plotted with the dotted line.
}\label{spe}
\end{figure}

The characteristics at the final time $t=t_f$ are shown in Table~\ref{DNS_characteristics}.
The Iroshnikov and Kraichnan (IK) microscale $\eta_{\mathrm{IK}}$ is defined by $ \left( \nu^{2} b_{0} / {\bar \epsilon}_t \right)^{1/3}$, where $b_{0}= \left( 2 {\bar E}_b / 3 \right)^{1/2}$.
The kinetic and magnetic Taylor microscale Reynolds numbers are given by 
$R^{u}_{\lambda} = u_0\lambda_{u}/\nu $ and $R^b_{\lambda}= b_0 \lambda_{b}/\eta$, respectively,
where $\lambda_{u}= (15 \nu u_{0}^{2} / {\bar \epsilon_\nu})^{1/2}$ and $\lambda_{b}=(15 \eta b_{0}^{2} /{\bar \epsilon}_{\eta})^{1/2}$. 
Figure \ref{spe} plots the kinetic energy spectrum $E_u(k)$ and the magnetic one $E_b(k)$ at $t=t_f$.
From these results, we can see that $E_b(k)$ is larger than $E_u(k)$ at any $k$ resolved in the DNS.

\section{Numerical Examination of Four-Fifths Law}
We examine the 4/5 law, using the DNS data at $R_\lambda^u = 158$ and $R_\lambda^b = 323 $ in the next subsection.
The examination is based on the generalized KHK equation (\ref{MHDKHu}).
Then, we provide some insight into the directional anisotropy of the third-order moments $6\langle b_L^2 \delta u_L \rangle ({\bm r})$ and $-\langle (\delta u_L)^3 \rangle ({\bm r})$ in Sec. IV B.
One may ask what we can say about inhomogeneity, but this is beyond the scope of this study.

\subsection{Examination based on generalized K\'arm\'an-Howarth-Kolmogorov equation}
\begin{figure}[tb]
\begin{center}
\includegraphics[width=8cm,keepaspectratio]{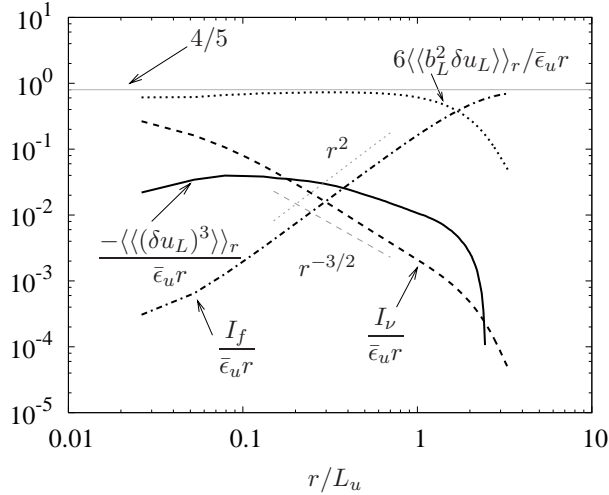}
\end{center}
\caption{
The $r/L_u$ dependence of the normalized forcing term $I_f/({\bar \epsilon}_u r)$ and viscous term $I_\nu/({\bar \epsilon}_u r)$ together with $-\langle \langle (\delta u_L)^3 \rangle \rangle_r/({\bar \epsilon}_u r)$ and $6\langle \langle b_L^2\delta u_L \rangle \rangle_r/({\bar \epsilon}_u r)$.
The gray solid line denotes $4/5$.
As references, the power laws, $r^{2}$ and $r^{-3/2}$, are plotted with the gray dotted and gray dashed lines, respectively.
}\label{forvis45}
\end{figure}

\begin{figure}[tb]
\begin{center}
\includegraphics[width=8cm,keepaspectratio]{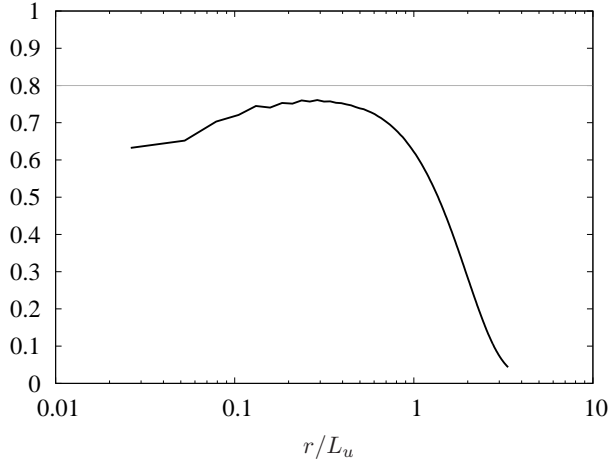}
\end{center}
\caption{Sum of normalized third-order terms, $\{-\langle \langle (\delta u_L)^3 \rangle \rangle_r+6\langle \langle b_L^2\delta u_L \rangle \rangle_r\}/({\bar \epsilon}_u r)$, denoted by the black solid curve, vs $r/L_u$. 
The gray solid line denotes $4/5$.
}\label{nonlin45}
\end{figure}

Figure \ref{forvis45} plots the DNS values of the third-order terms $6\langle \langle b_L^2 \delta u_L \rangle \rangle_r$ and $-\langle \langle (\delta u_L)^3 \rangle \rangle$ , both of which are normalized by ${\bar \epsilon}_u r$, as functions of $r/L_u$ on a log-log scale.
The normalized forcing and viscous terms, $I_f/({\bar \epsilon}_u r)$ and $I_\nu/({\bar \epsilon}_u r)$ are also plotted.
No DNS values at the smallest $r/L_u$ (i.e., $r/L_u \simeq 0.01$) are shown, 
because Simpson's rule is used in integrating ${\tilde r}^3 H_\alpha({\tilde r})$ $(\alpha=f,\nu,u,b)$ over ${\tilde r}$ in Eq. (\ref{Ialpha}). 
We observe that $6\langle \langle b_L^2\delta u_L\rangle \rangle_r$ is very dominant over $-\langle \langle (\delta u_L)^3 \rangle \rangle_r$ for all $r$.
This predominance is in accordance with the result obtained by Yousef {\it et al}. \cite{YRS}, at the scales except large scales.
They analyzed DNS data obtained in Ref. \cite{Scheko} showing that $E_b (k) < E_u(k)$ for small wave numbers $k$, i.e., large $r$, 
in contrast with the present DNS data.
The predominance in our results may be due to influence of the large-scale magnetic field.
It directly influences the term $\langle b_L^2\delta u_L\rangle$ which cannot be expressed by using only $\delta b_L$ and $\delta u_L$.
Because $6\langle \langle b_L^2\delta u_L\rangle \rangle_r$ and $- \langle \langle (\delta u_L)^3 \rangle \rangle_r$ respectively result from the nonlinear terms $b_j \partial_j b_i$ and $u_j \partial_j u_i$,
the 
predominance is also consistent with previous DNSs in Refs. \cite{Doma,Teaca}, where the flux for $b_j \partial_j b_i$ is dominant over that for $u_j \partial_j u_i$.
The departure of $6\langle \langle b_L^2\delta u_L\rangle \rangle_r/({\bar \epsilon}_u r)$ from $4/5$ remains small when $r$ decreases.
This small departure is also consistent with the result in Ref. \cite{YRS}.
It can be seen that $6\langle \langle b_L^2 \delta u_L \rangle \rangle_r/({\bar \epsilon}_u r)$ in Fig. \ref{forvis45} and the sum of the normalized third-order terms $\{ 6\langle \langle b_L^2\delta u_L\rangle \rangle_r  - \langle \langle (\delta u_L)^3 \rangle \rangle_r \}/({\bar \epsilon}_u r) $ in Fig. \ref{nonlin45} are almost constant in the range $ 0.2 \lesssim r/L_u \lesssim 0.5 $.
The range corresponds to the wave number range $0.06 \lesssim k \eta_{\mathrm{IK}} \lesssim 0.14$, where $k=\pi/r$.
The maximum value of the sum is about $0.76$ at $r/L_u \simeq 0.29$.
The normalized term $-\langle \langle (\delta u_L)^3 \rangle \rangle_r/({\bar \epsilon}_u r)$ is similar to $10^{-2}$, as shown in Fig. \ref{forvis45}.
It is comparable to $I_f/({\bar \epsilon}_u r)$ and $I_\nu/({\bar \epsilon}_u r)$ in the range $ 0.17 \lesssim r/L_u \lesssim 0.38$. 
Except for this range, $-\langle \langle (\delta u_L)^3 \rangle \rangle_r$ is smaller than either the forcing  term $I_f$ or the viscous term $I_\nu$.

\begin{figure}[tb]
\begin{center}
\includegraphics[width=8cm,keepaspectratio]{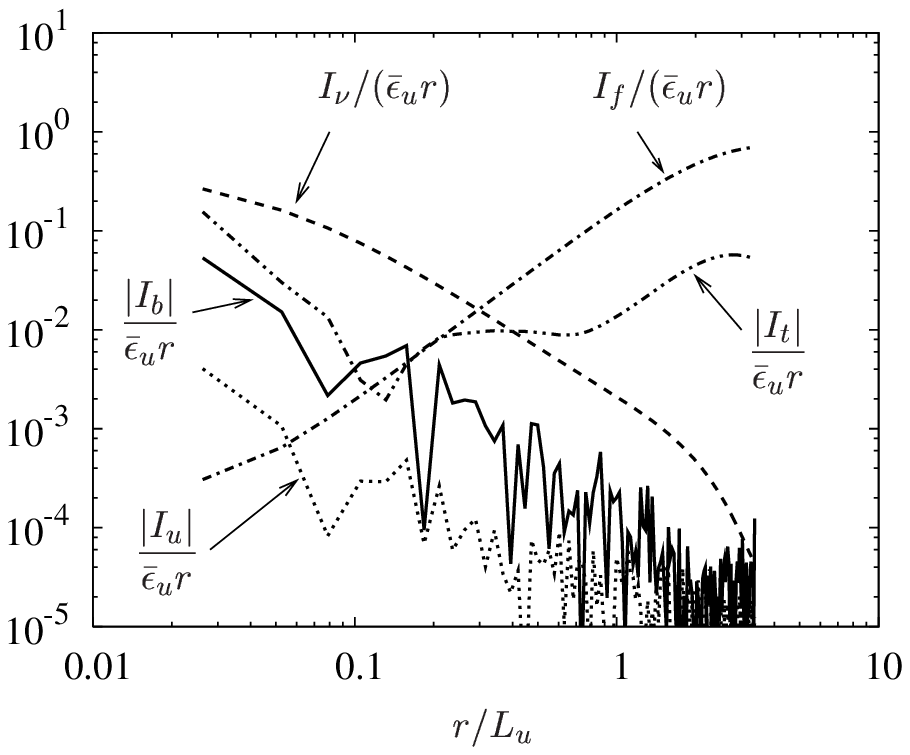}
\end{center}
\caption{ 
Magnitudes of the normalized anisotropic terms, $|I_u|/({\bar \epsilon}_u r)$ and $|I_b|/({\bar \epsilon}_u r)$, and the magnitude of the normalized nonstationary term, $|I_t|/{\bar \epsilon}^u r$, vs $r/L_u$, compared with the viscous and forcing terms, $I_\nu/({\bar \epsilon}_u r)$ and $I_f/({\bar \epsilon}_u r)$. 
}\label{ani_non}
\end{figure}

The influence of the forcing term becomes weaker rapidly, 
as the separation distance $r$ decreases.
The normalized forcing term $I_f/({\bar \epsilon}_u r)$ fits well to a simple scaling, $I_f/({\bar \epsilon}_u r) \propto r^2$, in the range $ 0.2 \lesssim r/L_u \lesssim 0.5 $.
This scaling agrees with previous theoretical predictions of the effect of forcing on the 4/5 law for HD turbulence \cite{Novikov,Fukayama,Gotoh,Kaneda2008} and DNS results investigating this effect \cite{Fukayama,Gotoh,Kaneda2008}. 
It can be seen that the normalized viscous term $I_\nu/({\bar \epsilon}_u r)$ increases with decreasing $r$ approximately as $I_\nu/({\bar \epsilon}_u r) \propto r^{-3/2}$ in the range $ 0.2 \lesssim r/L_u \lesssim 0.5 $.
The scaling $r^{-3/2}$ comes from the IK scaling in physical space (i.e., $ \langle D_{ii} \rangle_r \propto r^{1/2}$), 
which is equivalent to a $k^{-3/2}$ spectrum in wave number space.

Next, we discuss the influences of the anisotropic terms, $I_u$ and $I_b$, and the nonstationary term, $I_t$.
It should be recalled that these are never zero in real turbulence, 
even if all of the terms can be safely neglected.
In Fig. \ref{ani_non}, we find that $|I_u|/({\bar \epsilon}_u r)$ and $|I_b|/({\bar \epsilon}_u r)$ are at least one order of magnitude smaller than $I_\nu/({\bar \epsilon}_u r)$ at the scales except large scales.
This finding shows that the influence of these anisotropic terms on the 4/5 law is not significant compared to that of the viscosity, at least in our DNS.
It can be seen that $|I_u| \lesssim |I_b|$ for $r/L_u \lesssim 0.5$.
The influence of the anisotropy of $6 \langle \langle b_L^2 \delta u_L \rangle \rangle_r$ on the 4/5 law is stronger than that of $-\langle \langle (\delta u_L)^3 \rangle \rangle_r$.
This stronger influence is suggested to be due to the predominance of $6 \langle \langle b_L^2 \delta u_L \rangle \rangle_r$ over $-\langle \langle (\delta u_L)^3 \rangle \rangle_r$.
If we take into account that $-\langle \langle (\delta u_L)^3 \rangle \rangle_r$ is at most $O(10^{-2})$ and $6 \langle \langle b_L^2 \delta u_L \rangle \rangle_r$ is similar to $0.7$,
then we find $-|I_u|/\langle \langle (\delta u_L)^3 \rangle \rangle_r \sim |I_b|/(6 \langle \langle b_L^2 \delta u_L \rangle \rangle_r)$.
Therefore, it is shown that $-\langle \langle (\delta u_L)^3 \rangle \rangle_r$ is as anisotropic as $6 \langle \langle b_L^2 \delta u_L \rangle \rangle_r$ for $0.2 \lesssim r/L_u \lesssim 0.5$ in the sense studied here.

Figure \ref{ani_non} also plots the magnitude of the nonstationary term $|I_t|/({\bar \epsilon}_u r)$,
which was computed by the use of Eq. (\ref{MHDKHu}).
We can see that $|I_t|/({\bar \epsilon}_u r)$ is much smaller than $I_\nu/({\bar \epsilon}_u r)$ at small scales, and much smaller than $I_f/({\bar \epsilon}_u r)$ at large scales.
However, it is also observed that around $ r/L_u \simeq 0.3 $,
$|I_t|$ is not very much smaller than $I_\nu$  and $I_f$,
while $|I_t|$ is one order of magnitude larger than $|I_b|$.
Therefore, the departure of the maximum value of $\{ 6\langle \langle b_L^2\delta u_L\rangle \rangle_r  - \langle \langle (\delta u_L)^3 \rangle \rangle_r \}/({\bar \epsilon}_u r) $ from 4/5 is mainly due to the influences of forcing, viscosity, and non-stationarity.
It is noted that the influence of the nonstationary term is expected to become small, 
if one takes a time average of the statistics studied here for a long enough time interval.
\subsection{Directional anisotropy in third-order longitudinal moments}
In experiments and observations, 
it is easier to measure quantities using one particular direction of ${\bm r}$ [e.g. $\langle b_L^2 \delta u_L \rangle ({\bm r})$], 
rather than those averaged over a spherical shell, such as $\langle\langle b_L^2 \delta u_L \rangle \rangle_r (r)$.
Hence, one may approximate the latter as the former,
which raises the question of how accurate such an approximation is.

To get some ideas about the degree of the accuracy, we introduce 
\begin{eqnarray}
\Delta_i^u (r) &=& -\langle (\delta u_L)^3 \rangle(r{\bm e}_i) + \langle \langle (\delta u_L^3    \rangle({\bm r})) \rangle_r ,\\
\Delta_i^b (r) &=& 6 \left\{ \langle b_L^2 \delta u_L \rangle(r{\bm e}_i) - \langle \langle b_L^2 \delta u_L   \rangle ({\bm r}) \rangle_r \right\},
\end{eqnarray}
where ${\bm e}_i$ is the unit vector in the $i$th direction [i.e., ${\bm e}_1=(1,0,0)$, ${\bm e}_2=(0,1,0)$ and ${\bm e}_3=(0,0,1)$].
We also define their averages over $i=1,2,3$ as
\begin{equation}
\Delta^\alpha_{\mathrm{ave}} (r) = \frac{1}{3} \sum_{i=1}^3 \Delta_i^\alpha (r) \quad (\alpha=u,b) .\end{equation}
If flow is isotropic in a strict sense, 
then $\Delta_i^\alpha(r) =0 $ for any $i$, and thus $\Delta_{\mathrm{ave}}^\alpha (r) =0$.
Therefore, $\Delta_i^\alpha(r)$ and $\Delta_{\mathrm{ave}}^\alpha (r) $ are measures of the strength of the directional anisotropy.

\begin{figure}[t]
\begin{center}
\includegraphics[width=8cm,keepaspectratio]{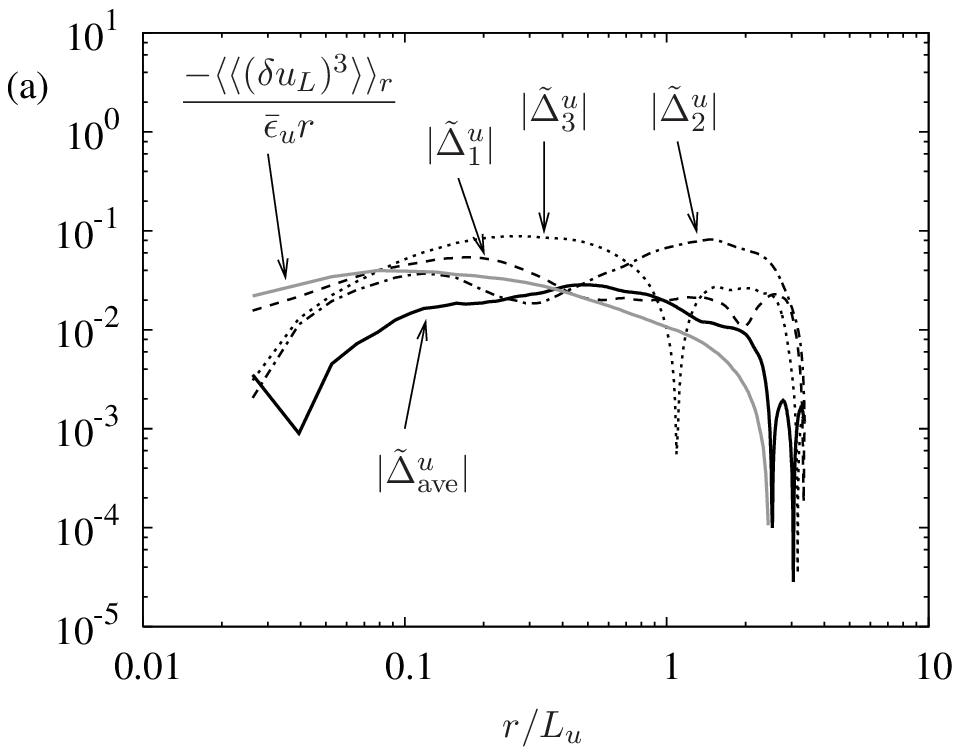}
\includegraphics[width=8cm,keepaspectratio]{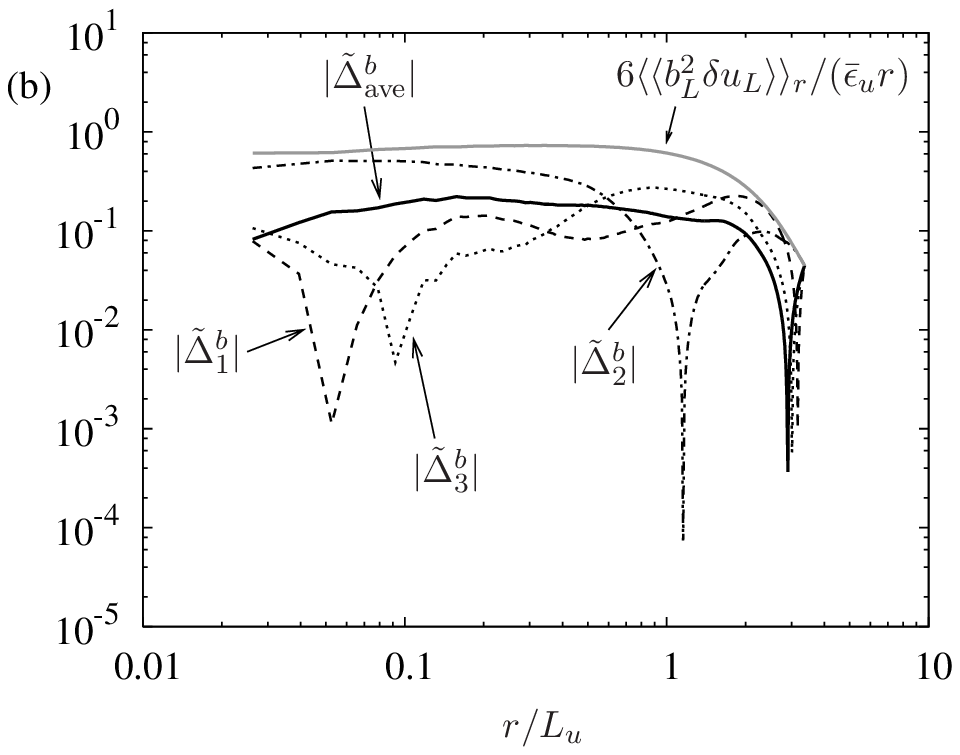}
\end{center}
\caption{
(a) Normalized anisotropic measures $|{\tilde \Delta^u_i}(r)|$ $(i=1,2,3)$ and the average $|{\tilde \Delta^u_{\mathrm{ave}}}(r)|$ vs $r/L_u$.
(b) Measures $|{\tilde \Delta^b_i}(r)|$ $(i=1,2,3)$ and $|{\tilde \Delta^b_{\mathrm{ave}}}(r)|$ vs $r/L_u$.
The terms $-\langle \langle (\delta u_L)^3 \rangle \rangle_r/({\bar \epsilon}_u r)$ and $6\langle \langle b_L^2\delta u_L \rangle \rangle_r/({\bar \epsilon}_u r)$ are plotted in (a) and (b), respectively.
}\label{dirani}
\end{figure}

Figure \ref{dirani} plots the magnitude of the anisotropic measures normalized by ${\bar \epsilon}_u r$, denoted by ${\tilde \Delta_i^\alpha}$ and ${\tilde \Delta_{\mathrm{ave}}^\alpha}$, 
where ${\tilde \Delta_i^\alpha} = \Delta_i^\alpha/({\bar \epsilon_u} r)$, 
and ${\tilde \Delta_{\mathrm{ave}}^\alpha} = \Delta_{\mathrm{ave}}^\alpha/({\bar \epsilon_u} r)$.
For comparison, $ -\langle\langle (\delta u_L)^3 \rangle\rangle_r/({\bar \epsilon}_u r)$ and $6 \langle\langle  b_L^2 \delta u_L \rangle\rangle_r/({\bar \epsilon}_u r)$ are also presented. 
We have the following four observations in the range $ 0.2 \lesssim r/L_u \lesssim 0.5 $:
\begin{enumerate}
\item[(i)] The anisotropic measure $|{\tilde \Delta}_{\mathrm{ave}}^b |$ is one order of magnitude larger than $|{\tilde \Delta}_{\mathrm{ave}}^u |$, as expected.
\item[(ii)] The measure $|{\tilde \Delta}_{\mathrm{ave}}^u|$ is as large as $-\langle\langle (\delta u_L)^3 \rangle\rangle_r/({\bar \epsilon}_u r)$, 
while $|{\tilde \Delta}_{\mathrm{ave}}^b|$ is not very small, 
compared to $6\langle \langle b_L^2 \delta u_L \rangle \rangle_r/({\bar \epsilon}_u r)$.
\item[(iii)] Among ${\tilde \Delta}_i^u$ for $i=1,2,3$, ${\tilde \Delta}_3^u$ is the largest.
In contrast, ${\tilde \Delta}_2^b$ is dominant over ${\tilde \Delta}_1^b$ and ${\tilde \Delta}_3^b$.
\item[(iv)]
Comparison of Figs. \ref{forvis45} and \ref{dirani} shows that
$|{\tilde \Delta}_{\mathrm{ave}}^u|$ is comparable to $I_f/({\bar \epsilon}_u r)$, while $|{\tilde \Delta}_{\mathrm{ave}}^b|$ exhibits much larger values compared to $I_f/({\bar \epsilon}_u r)$.
\end{enumerate}
The result about $|{\tilde \Delta}_{\mathrm{ave}}^u|$ in (ii) is in contrast to the result in the case of HD turbulence \cite{Kaneda2008},
where the anisotropic measure of $-\langle (\delta u_L)^3 \rangle (r{\bm e}_i)$ averaged over $i$ was shown to be much smaller than $-\langle\langle (\delta u_L)^3 \rangle \rangle_r$.
For 3D MHD turbulence, it is shown in Refs. \cite{scaledependent,da1,da2,Pandit} that,
although the local dynamic alignment (i.e., local alignment or local anti-alignment between velocity and magnetic fields)
is pronounced 
the velocity does not align with the magnetic field globally.
One can speculate that this lack of the global dynamic alignment leads to result (iii); 
namely, the difference between the directions where $-\langle (\delta u_L)^3 \rangle ({\bm r})$ and $\langle b_L^2 \delta u_L \rangle ({\bm r})$ are the most anisotropic.
It is noted that $\Delta_i^u = 6\{- \langle u_L^2 u_L' \rangle (r{\bm e}_i)  +\langle \langle u_L^2 u_L' \rangle \rangle_r \}$ and $\Delta_i^b = 6\{ \langle b_L^2 u_L' \rangle (r{\bm e}_i)  - \langle \langle b_L^2 u_L' \rangle \rangle_r \}$.
Result (iv) suggests that the sensitivity of the anisotropy at small scales to the large-scale conditions is stronger than that of the forcing term, as was found in Ref. \cite{Kaneda2008}.
It is found that
the approximation of a spherical average by the average over only the three directions does not work very well, at least in the case studied here.
In contrast, for HD turbulence it has been shown that the average of $-\langle (\delta u_L)^3 \rangle ({\bm r})$ over three Cartesian directions is a good approximation of $-\langle\langle (\delta u_L)^3 \rangle \rangle_r$ \cite{Kaneda2008}.
Hence, these results suggest that the directional anisotropy has a substantial influence on the 4/5 law for MHD turbulence, compared to the case of HD turbulence.

\section{Conclusions}
We have examined the 4/5 law for third-order moments in 3D incompressible MHD turbulence in a periodic box in the absence of a uniformly imposed magnetic field.
The examination was based on a generalization of the KHK equation for isotropic MHD turbulence to anisotropic MHD turbulence.
The generalization is made by means of a spherical average over the direction of the separation vector.
The influences of the viscous force, the external force, anisotropy, and non-stationarity in the generalized equation were quantified.
We analyzed instantaneous DNS data on 3D forced incompressible MHD turbulence in a periodic box at the moderate kinetic and magnetic Taylor microscale Reynolds numbers, $R_\lambda^u=158$ and $R_\lambda^b =323$.
The magnetic Prandtl number is set to one and the number of grid points is $512^3$.

It was confirmed that 
$6\langle \langle b_L^2 \delta u_L \rangle ({\bm r}) \rangle_r$ is dominant over $-\langle \langle (\delta u_L)^3 \rangle ({\bm r}) \rangle_r$.
The sum of the normalized third-order moments $\{6\langle \langle b_L^2 \delta u_L \rangle \rangle_r- \langle \langle (\delta u_L)^3 \rangle \rangle_r\}/({\bar \epsilon}_u r) $ is almost constant in the range $0.2 \lesssim r/L_u \lesssim 0.5$.
In other words, the sum obeys the linear scaling with $r$ well.
We have found that 
the small departure of the sum from 4/5 in the range is mainly due to the influences of the viscous, forcing, and nonstationary terms in the generalized equation.
The influence of the anisotropic terms is shown to be negligible compared to that of the viscous term at small scales.
However, the dependences of $6\langle b_L^2 \delta u_L \rangle ({\bm r})$ and $-\langle (\delta u_L)^3 \rangle ({\bm r})$ on the direction ${\bm r}$ are so strong that 
the directional anisotropy of $6\langle b_L^2 \delta u_L \rangle $ and $-\langle (\delta u_L)^3 \rangle$ has substantial influence on the 4/5 law, at least in the case studied here.
Averaging of $6\langle b_L^2 \delta u_L \rangle $ and $-\langle (\delta u_L)^3 \rangle$ in terms of ${\bm r}$ over three Cartesian directions does not provide good approximations of the spherically averaged quantities, $6\langle \langle b_L^2 \delta u_L \rangle ({\bm r}) \rangle_r$ and $-\langle \langle (\delta u_L)^3 \rangle ({\bm r}) \rangle_r$.
This is in contrast to homogeneous quasi-isotropic HD turbulence, 
where averaging $-\langle (\delta u_L)^3 \rangle$ over the three directions is a good approximation of $-\langle \langle (\delta u_L)^3 \rangle \rangle_r$ \cite{Kaneda2008}.
It is conjectured that time averaging over more large-eddy turn over times may be required for MHD turbulence, compared to HD turbulence, 
before the averaging over three Cartesian directions can be a good approximation of the spherical averaging.

Examination of the exact law for the third-order terms arising from the induction equations (\ref{induction}) remains an issue.
However, this is beyond the scope of this study.
The formulations of the third-order moments, which was derived in Ref. \cite{Chandra} under the assumption of flow isotropy, are not so simple, 
and are not expressed only by the longitudinal components of velocity and magnetic fields
[see Eqs. (69) and (105) in Ref. \cite{Chandra}].
These moments are obtained from $\langle u_i b_j b_k' \rangle-\langle u_j b_i b_k' \rangle$, 
which is antisymmetric in the indices $i$ and $j$.
Using the Els\"asser variables, $z^{\pm}_i=u_i \pm b_i$,
we have $\langle u_i b_j b_k' \rangle$ and $-\langle u_j b_i b_k' \rangle$ in $\langle z_i^\pm ({\bm x}) z_j^\mp ({\bm x}) z_k^\pm ({\bm x}') \rangle$, 
which is not symmetric in the indices $i$ and $j$. 
In future work, it would be interesting to examine the dependence of the statistics studied here on the Reynolds numbers and scales, on the basis of DNS data on higher Reynolds number MHD turbulence.
Examination of the exact statistical laws for kinetic energy, magnetic energy, cross helicity, and magnetic helicity for various types of MHD turbulence 
(e.g., 3D homogeneous anisotropic MHD turbulence in the presence of an imposed mean magnetic field and two-dimensional homogeneous MHD turbulence)
is expected to help provide a quantitative understanding of the universality.
As discussed in Ref. \cite{Galtier}, in the examination of such anisotropic MHD turbulence, it is important to pay attention to what average is taken in terms of ${\bm r}$.

\noindent {\bf Acknowledgments}: \\
The computations were carried out on the FX1 and M9000 systems at the Information Technology Center of Nagoya University.
This work was supported by a Grant-in-Aid for Young Scientists (B) 22740255 from the Ministry of Education, Culture, Sports, Science and Technology and by a Grant-in-Aid for Scientific Research (S) 20224013 from the Japan Society for the Promotion of Science.
The author is grateful to H. Hagiwara and Y. Kondo for their numerical supports,
and also acknowledges Professor Y. Kaneda at Nagoya University for helpful discussion on the four-fifths law for incompressible HD turbulence.


\end{document}